\documentstyle[prb,multicol,aps,epsfig]{revtex}

\begin{document}

\title{Local polariton modes and resonant tunneling of electromagnetic waves through periodic Bragg multiple quantum well structures}

\author{Lev I. Deych\dag, Alexey Yamilov\ddag, and A. A. Lisyansky\ddag}
\address{{\dag} Department of Physics, Seton Hall University, South Orange, NJ 07079\\
{\ddag} Department of Physics, Queens College of CUNY, Flushing, NY}
\date{\today}
\maketitle

\begin{abstract}
 
We study analytically defect polariton states in Bragg multiple-quantum-well structures and defect induced changes in transmission and reflection spectra. Defect layers can differ from the host layers in three ways: exciton-light coupling strength, exciton resonance frequency, and inter-well spacing. We show that a single defect leads to two local polariton modes in the photonic bandgap. These modes cause peculiarities in reflection and transmission spectra. Each type of defect can be reproduced experimentally, and we show that each of these plays a distinct role in the optical properties of the system. For some defects, we predict a narrow transmission window in the forbidden gap at the frequency set by parameters of the defect. We obtain analytical expressions for corresponding local frequencies as well as for reflection and transmission coefficients. We show that the presence of the defects leads to resonant tunneling of the electromagnetic waves via local polariton modes accompanied by resonant enhancement of the field inside the sample, even when a realistic absorption is taken into account. On the basis of the results obtained, we make recommendations regarding the experimental observation of the effects studied in readily available samples.

\end{abstract}


\section{Introduction}

Optical properties of multiple quantum wells (MQW) have attracted a great deal of interest recently.\cite{Citrin,Ivchenko,Andreani,Bjork,Hubner,Stroucken,Vladimirova,Khitrova} Unlike other types of superlattices, excitons in MQW are confined in the planes of the respective wells, which are separated by relatively thick barriers. Therefore, the only  coupling between different wells is provided by the radiative optical field. The coupling results in MQW polaritons -- coherently coupled quasi-stationary excitations of quantum well (QW) excitons and transverse electromagnetic field. The spectrum of short MQW structures consists of a number of quasi-stationary (radiative) modes with finite life-times. This spectrum is conveniently described in terms of super- or sub-radiant modes\cite {Citrin,Andreani,Bjork}. When the number of wells in the structure grows, the life-time of the former decreases, and the life-time of the latter increases. In longer MQW structures, however, this approach becomes misleading, as discussed in Ref.\onlinecite{DeychQW}, and a more appropriate description is obtained in terms of stationary modes of an infinite periodic structure. The spectrum of MQW polaritons in this case consists of two branches separated by a gap with a width proportional to the exciton-light coupling constant $\Gamma. $\cite{Citrin,Andreani} The length at which this description becomes preferable, usually depends upon a problem at hand. For instance, even though the discrete structure of the sub-radiant modes of 100 long MQW in Refs.\onlinecite{Hubner,Khitrova} could be resolved, the description of the gap region in terms of the super-radiant mode leads to an apparent contradiction with the absence of significant luminescence  in this region. If one is interested in properties of the gap region, a ``long system" usually means that it is longer than the penetration length of radiation into the sample. The latter length depends upon the frequency, therefore the system can be long enough for frequencies close to the center of the gap, and still ``short" for frequencies in the vicinity of the band edges.  Systems similar to those studied in Ref.\onlinecite{Khitrova} can be considered sufficiently long for $90\%$ of the gap region when the number of wells becomes close to $200$.

In a number of papers \cite {Ivchenko,Hubner,Vladimirova,Khitrova} it was shown that the width of the polariton bandgap can be significantly increased by tuning the interwell spacing, $a$, to the Bragg condition, $a=\lambda_0 /2$, where $\lambda_0$ is the wavelength of the light at the exciton frequency $\Omega _0$. Under this condition, boundaries of two adjacent gaps become degenerate, and one wider gap with the width proportional to $\sqrt{\Gamma }$ is formed. Detuning of the lattice constant from the exact Bragg condition removes the degeneracy and gives rise to a conduction band in the center of the Bragg gap.\cite{DeychQW} A well-pronounced Bragg polariton gap was observed in recent experiments\cite{Khitrova} with $InGaAs/GaAs$ Bragg structures with the number of wells up to $100$. These experiments convincingly  demonstrate that despite homogeneous and inhomogeneous broadening, the coherent exciton-photon coupling  in long MQW is experimentally feasible. Polariton effects arising as a result of this coupling open up new opportunities for manipulating optical properties of quantum heterostructures. 

One such opportunity is associated with introducing defects in MQW structures. These defects can be either QW's of different compositions replacing one or several ``host" wells, or locally altered spacing between elements of the structure. It is well known in the physics of regular crystals (see, for instance, Ref. \onlinecite{Lifshitz}) that local violations of otherwise periodic structures can lead to the appearance of local modes with frequencies within spectral gaps of host structures.  This idea was first applied to MQW in Ref.\onlinecite{Citrinlocal}, where it was shown that different defects can indeed give rise to local exciton-polariton modes in infinite MQW. Unlike regular local modes in 3-$d$ periodic structures, these modes are localized only in the growth direction of the MQW, while they can propagate along the planes of the wells. Therefore, one should clearly distinguish these defect polariton modes from well known interface modes in layered systems or non-radiative two-dimensional polariton modes in ideal MQW. \cite{Dereux,Lahlaouti,Andreanireview} The latter exist   only with the in-plane wave numbers, $k_{||}$, exceeding certain critical values, while the local mode in a defect MQW structure exists at  $k_{||}=0$  and can be excited even at normal incidence.

In this paper we present results of detailed studies of local polariton modes (LPM's) produced by four different types of individual defects in Bragg MQW structures. The peculiar structure of Bragg MQW's results in a wider than usual polariton gap, which is actually formed by two gaps with degenerate boundaries. This property has a profound effect on the properties of local modes, leading, for instance, to emergence of two local states from a single defect. The similar effects of doubling the number of local modes was also predicted in the case of braggoritons - excitations arising inside photonic band  gaps of periodic structures made of a dispersive material, if the resonance frequency of the dielectric functions happens to belong to the gap.\cite{Raikh}

We neglect in-plane disorder in individual QW's, and assume that apart from a deliberately introduced single defect, the structure remains ideally periodic. We consider LPM's with zero in-plane wave vector only. Such modes can be excited by light incident in the growth direction of the structure, and can result in the resonance transmission of light with gap frequencies. This effect is studied both analytically and numerically with homogeneous broadening taken into account phenomenologically. Equations describing dynamics of the light-exciton interaction in this situation are essentially equivalent to a model of the one-dimensional chain of dipoles used in the series of our previous works where LPM's in polar crystals were discussed. \cite{Deych,PRBlocal,JOSAB} Similar equations also appear in the theory of atomic optical lattices.\cite{Deutsch} The essential difference between results presented in this paper and previous studies stems from the peculiarities of the Bragg arrangement. Using Greens' function and transfer matrix formalisms, we study both eigenfrequencies of LPM's for different types of defects and transmission properties of the defect structures. Taking into account homogeneous broadening and using parameters of the system studied experimentally in Refs.\onlinecite{Hubner,Khitrova}, we predict which defects will produce the most significant changes in transmission and reflection properties of realistic MQW structures.


\section{Defect modes in Bragg MQW}

In order to describe optical properties of QW's one has to take into account the coupling between retarded electromagnetic waves and excitons. This is usually done with the use of the non-local susceptibility determined by energies and wave functions of a QW exciton.\cite{Citrin,Andreanireview} The treatment of the exciton subsystem can be significantly simplified if the interwell spacing is much larger than the size of a well itself. In $InGaAs/GaAs$ MQW structures studied in Ref.\onlinecite{Khitrova}, on which we base our numerical examples, the width of the QW layer amounts only to about 10\% of the period of the structure. In this case, one can neglect the overlap of the exciton wave functions from neighboring wells and  assume that an interaction between well excitons occurs only due to coupling to the light. It is also important that the width of the wells is also considerably smaller than the exciton's Bohr radius, and, therefore, one can neglect spatial extent of the wells, and describe them with polarization density of the form: ${\bf P}({\bf r},z)={\bf P}_n({\bf r})\delta(z-z_n)$, where ${\bf r}$ is an in-plane position vector,  $z_n$ represents a coordinate of the $n$-th well, and ${\bf P}_n$ is a surface polarization density of the respective well. Optical response of the system even in the presence of inhomogeneous broadening is  successfully described by the linear dispersion theory (see, for instance, Ref.\onlinecite{LDT}) based upon a single oscillator phenomenological expression for the exciton susceptibility 
\begin{equation}
\chi= \frac{d^2}{\Omega_0^2-\omega^2-2i\gamma_{hom}\omega},
\label{susc}
\end{equation} 
where $\Omega_0$ and $\gamma_{hom}$ are $1s$ exciton frequency and relaxation parameter respectively, $d$ is a parameter describing the light-exciton coupling.  For our purposes, however, it is more convienient to consider explicitely equations of exciton dynamics coupled with Maxwell equations for the electromagnetic field. In the case of waves propagating along the growth direction of the structure, which is considered in the present paper, only excitons with the zero in-plane wave number are coupled to light, and the equations of motion can be presented in the form:
\begin{equation}
\left( \Omega _{n}^{2}-\omega ^{2}-2i\gamma_{hom}\omega\right) P_{n}=\frac{1}{\pi }c\Gamma _{n} E(z_{n}),
\label{P}
\end{equation}
\begin{equation}
\frac{\omega ^{2}}{c^{2}}E\left( z\right) +\frac{d^{2}E\left( z\right) }{ dz^{2}}=-4\pi \frac{\omega ^{2}} {c^{2}} \sum\limits_{n} P_{n} \delta \left( z-z_{n}\right),
\label{E}
\end{equation}
where $\Omega _{n}$ is the exciton frequency of the $n$-th QW. The strength of the exciton-photon coupling in Eq. (\ref{P}) is described by the parameter $\Gamma _{n}$ equal to the radiative decay rate of a single $n$-th well. This parameter is related to the parameter $d$ of Eq. (\ref{susc}) according to $d^2=(c\Gamma)/(\pi w)$, where $w$  characterizes the spatial extent of exciton wave functions in the growth direction (we assumed it to be zero when we described the exciton polarization by the $\delta$-function). In what follows, we refer to $\Gamma_n$ as to a coupling parameter. The relation of $\Gamma_n$ to the radiative life time can be established, for instance, if one uses Eqs. (\ref{P}) and (\ref{E})  to obtain the reflection coefficient for a single $n$-th QW, $r_n$, 
\begin{equation}
r_n=\frac{2i\omega\Gamma_n}{\Omega_0^2-\omega^2-2i\omega(\gamma_{hom}+\Gamma_n)}\approx \frac{i\Gamma_n}{\Omega_0-\omega-i(\gamma_{hom}+\Gamma_n)}
\label{rn}
\end{equation}
which in the resonance approximation (the last expression in Eq. (\ref{rn})) coincides with the standard linear dispersion theory expression used in Refs. \onlinecite{Ivchenko,Khitrova} and others. 

In an infinite pure system, all $\Gamma _{n}=\Gamma _{0},\ \Omega _{n}=\Omega _{0},\ z_{n}=na$, where $a=\lambda_0/2$ is the Bragg's interwell separation. Eq. (\ref{E}) describes an electromagnetic wave of one of the degenerate transverse polarizations, propagating in the growth direction of the structure. This equation
coincides with equations used in Refs. \onlinecite {Deych,PRBlocal,JOSAB,Deutsch} for  one-dimensional chains of atoms.  
The spectrum of ideal periodic MQW's with homogeneous broadening parameter $\gamma_{hom}=0$ has been studied in many papers.\cite{Citrin,Andreani,DeychQW,Deutsch,Keldysh,Ivchenko2} In the specific case of Bragg structures, the exciton resonance  frequency, $\Omega _0$, is at the center of the  bandgap determined by the inequality  $\omega _{l} <\omega <\omega _{u}$, where $\omega _{l}=\Omega _{0}\left( 1-\sqrt{2\Gamma _{0} / \pi\Omega _0}\right)$ and $\omega _{u}=\Omega _{0}\left( 1+\sqrt{2\Gamma _{0} / \pi\Omega _0}\right)$.\cite{DeychQW} In the formed bandgap the electromagnetic waves decay with the penetration (localization) length,
\begin{equation}
l_{loc}^{-1}=a^{-1} \ln \left( 
\cos \left( k_0 a \right) +\frac{\beta}{2} \sin \left( k_0 a\right)  -
\sqrt{\left[ \cos \left( k_0 a \right) +\frac{\beta}{2} \sin \left( k_0 a\right) \right]^2-1}
\right),  
\label{extinction_length}
\end{equation}
where $k_0=\omega /c$ and 
\begin{equation}
\beta =\frac{4\Gamma _{0}\omega }{\omega ^{2}-\Omega _{0}^{2}}.
\label{polarizability}
\end{equation}
Parameter $\beta$ is proportional to the susceptibility of the well in the absence of exciton relaxation, Eq. (\ref{susc}), and is the main quantity used in our theory to characterize the optical response of the single well. The homogeneous broadening can be taken into account by adding $2i\gamma_{hom}\omega $ in the denominator of Eq. (\ref{polarizability}); we, however, neglect it until the last section of the paper, where transmission and reflection spectra of defect MQW structures are discussed.

As it was mentioned earlier, for  structures, which are longer than the penetration length, $l_{loc}$, for the most of the band-gap, consideration based upon modes of the infinite periodic structure is more appropriate. Using Eq. (\ref{extinction_length}) one can show that in the case of 140 (for GaAs/AlGaAs) or 200 (for InGaAs/GaAs) wells long  MQW's of the type considered in Refs.\onlinecite{Hubner,Khitrova}, $l_{loc}>Na$ for $90\%$ of the bandgap of the infinite structure characterized by the boundaries $\omega_l$ and $\omega_p$. This bandgap is the frequency region where we look for new local states associated with the defects. In this paper, we consider four types of such defects. First of all, one can replace an original QW with a QW with different exciton frequency ($\Omega $-defect). This can be experimentally achieved by varying the composition of the semiconductor in the well. Another possibility is to change the coupling constant ($\Gamma$-defect) at one of the wells. Even though an experimental realization of this defect in its pure form is not straightforward, it is still methodologically important to consider such an idealized situation in order to be able to estimate how changes in $\Gamma$ could effect optical properties of the systems with real defects.  The third possibility is to perturb  an interwell spacing between two wells. Here we distinguish two defects, which we call $a$- and $b$-defects. The former is realized when the interwell spacing  between a pair of wells is changed (Fig. 1a). It can be seen that this induces a shift in the position of all wells that follow the defect, making it significantly nonlocal. The $b$-defect is produced when one shifts just one well keeping positions of the rest unchanged (Fig. 1b). Experimental realizations of these two defects is simple and can be done at the sample growth stage. In the following section we show that each of these types affects the optical properties of the MQW lattice in remarkably different ways.

\begin{figure}
\centering
\vspace{-0.0in}
\epsfxsize=3in \epsfbox{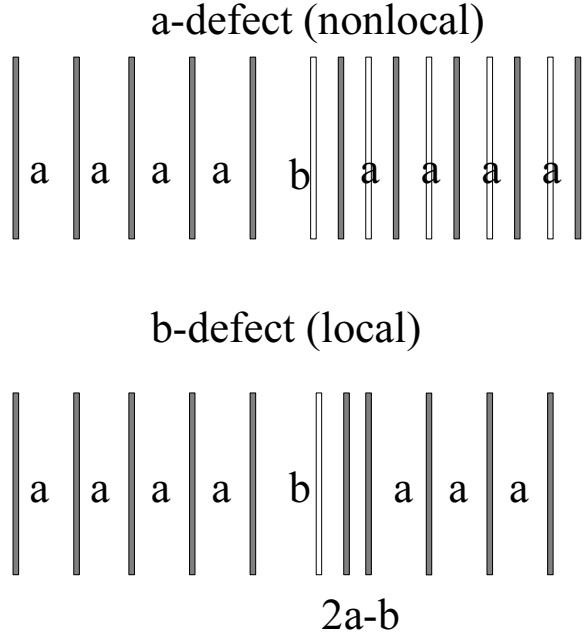}
\vspace{0.2in}
\caption{Two types of interwell spacing defects. The nonlocal $a$-defect (a) as opposed to the local $b$-defect (b). Solid bars represent locations of QW's in the defect lattice. The empty bars represent what would have been a perfectly ordered MQW lattice.}
\end{figure}

We start with $\Omega$- and $\Gamma$-defects which are similar in a sense that they both introduce perturbations in the equation of motion, Eq. (\ref{P}), which are localized at one site (the diagonal disorder). Therefore, they can be studied by the usual Green's function technique like the one used to treat the localized phonon states in impure crystals (see, for instance, Ref.\onlinecite{Lifshitz}). Using the  polarization Green's function defined in a standard way by adding $\delta_{n,n_0}$ to Eq. (\ref{P}), we can express the polarization of the $n$-th QW in terms of the polarization of the defect QW, $P_{n_0}$ as
\[
P_n=G(n-n_0) P_{n_0}.
\]
(see details of the derivation, as well as an expression for $G(n-n_0)$ in Ref. \onlinecite{PRBlocal}). Allowing $n=n_0$, and using expression for the Green function from Ref.\onlinecite{PRBlocal} one obtains the equation for  eigenfrequencies of LPM:
\begin{equation}
F_ {_{\Omega,\Gamma}}=\frac{\frac{\beta}{2} \sin \left( k_0 a\right) }{\sqrt{\left[ \cos \left( k_0 a \right) +\frac{\beta}{2} \sin \left( k_0 a\right) \right]^2-1}} ,  
\label{cond_inf_Om}
\end{equation}
where the function $F_{_{\Omega}}=\left(\Omega _{1}^{2}-\omega ^{2}\right)/\left(\Omega _{1}^{2}-\Omega _{0}^{2}\right)$ corresponds to the $\Omega$-defect and  the  function  $F_{_{\Gamma}}=\Gamma_{0}/\left(\Gamma_{1}-\Gamma_{0}\right)$ describes the $\Gamma$-defect; $\Omega_1$ and $\Gamma_1$ denote respective parameters of the defect layer. Eq. (\ref{cond_inf_Om}) is an exact consequence of Eqs. (\ref{P}) and (\ref{E}), and was first derived in Ref.\onlinecite{Deych} for the $\Omega$-defect. It  has been studied in that paper in the longwave approximation; it was found that the equation had one real valued solution for any $\Omega_1>\Omega_0$. We find that in the case of the Bragg structures, there are always two solutions for both types of defects, one below $\Omega_0$ and one above. This is a manifestation of the fact that the bandgap consists of two gaps positioned one right after the other. The above equations can be solved approximately when $\Gamma _{0} \ll \Omega_0$, which is the case for most materials. For the $\Omega$-defect, one solution demonstrates a radiative shift from the defect frequency $\Omega_1$, 
\begin{equation}
\omega _{def}^{(1)}=\Omega _{1}-\Gamma _{0}\frac{\Omega _{1}-\Omega _{0}}{\sqrt{\left( \omega _{u}-\Omega _{1}\right) \left( \Omega _{1}-\omega _{l}\right) }},        
\label{cond_inf_Om_app}
\end{equation}
while the second solution  splits off the upper or lower boundary depending upon the sign of $\Omega _{1}-\Omega_{0}$:  
\begin{equation}
\omega_{def}^{(2)}=\omega _{u,l} \pm \frac{1}{2} (\omega _{u}-\omega _{l})\left( \frac{\pi }{2}\frac{\Omega _{1}-\Omega _{0}}{\Omega _{0}}\right) ^{2},
\label{def2freq}
\end{equation}
where one chooses $\omega_u$ and ``$-$'' for $\Omega _{1}<\Omega_{0}$, and $\omega_l$ and ``$+$'' in the opposite case. This is illustrated in Fig. 2a. It can be seen that the shift of $\omega _{def}^{(1)}$ from the defect exciton frequency $\Omega _{1}$ is negative for $\Omega _{1}>\Omega_{0}$ and positive for $\Omega _{1}<\Omega_{0}$. The magnitude of the shift is of the order of the coupling constant $\Gamma _0$, which is usually rather small, and this fact, as it is shown in the next section, is crucial for the optical properties of the defect. The second local mode $\omega_{def}^{(2)}$ lies very close to the edges of the bandgap, and it would be, therefore, very difficult to distinguish it from the modes making up the allowed bands even for negligible dissipation.

\begin{figure}
\centering
\vspace{-0.0in}
\epsfxsize=3in \epsfbox{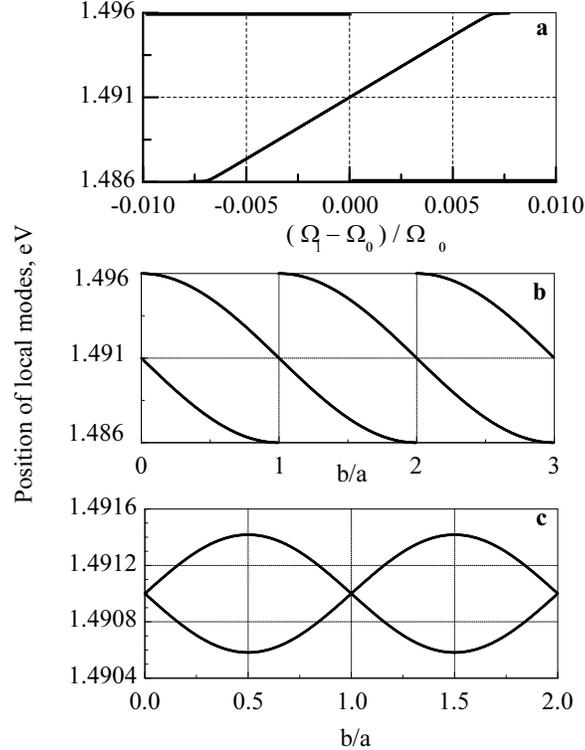}
\vspace{0.2in}
\caption{Positions of the local modes in the bandgap (1.486 - 1.496 eV) for the $\Omega $-defect (a), $a$-defect (b), and $b$-defect (c) as functions of the defect strengths. Numerical values of the exciton resonant frequency and the exciton-light coupling constant were taken the same as in $InGaAs/GaAs$ structures studied in Ref. 8.}
\end{figure}

In the case of the $\Gamma$-defect, one again finds two solutions of Eq. (\ref{cond_inf_Om}): 
\begin{equation}
\omega _{def}^{(1,2)}=\omega _{u,l} \pm 2\left(\Gamma_1-\Gamma_0\right)^{2}\left(\omega _{u}-\omega _{l}\right).
\label{Gamma_def}
\end{equation}
which exist only for $0<\Gamma _{1}<\Gamma _{0}$ and are  very close to the gap boundaries. The situation is similar to the second solution for the $\Omega$-defect, and one can conclude, therefore, that the $\Gamma$-defect would not affect significantly the optical spectra of the system. 

The next defect we consider, the $a$-defect, is shown in Fig. 1a.  One can see that this defect differs from the other two in a fundamental way. An increase in an interwell distance between any two wells automatically changes the coordinates of an infinite number of wells: $z_{n}=na$ for $n\leq n_{d}$ and $z_{n}=(b-a)+na$ for $n_{d}<n$, where $b$ is the distance between the $n$-th and $(n+1)$-th wells. Therefore, this defect is non-local and cannot be solved by the same methods as the two previous cases. The best approach to this situation is to consider the system of a finite length, $L$, and  match solutions for $n<n_d$ and $n>n_d+1$ with a solution for $na<z<na+(b-a)$, which is  schematically shown in Fig. 3. Eigen frequencies can be derived then considering limit $L\rightarrow \infty$. Solutions for the finite system can be constructed using the transfer matrix approach. The state of the system at the $m$-th well is described by a two dimensional vector  $v_{m}$ with components $E(x_{m})$ and $(1/k_0)(dE(x_{m})/dx)$:
\begin{equation}
v_{m}=\prod_{n=1} ^m \widehat{\tau}_{n}v_{0}=\widehat{T}v_{0}.
\label{transfer}
\end{equation}
The $2 \times 2$ transfer matrix $\widehat{\tau}_n$ at the $n$-th well is 
\begin{equation}
\widehat{\tau} _n=
\pmatrix{
\cos (k_0 a_n)+\beta \sin (k_0 a_n) & \sin (k_0 a_n) \cr 
-\sin (k_0 a_n)+\beta \cos (k_0 a_n) & \cos (k_0 a_n)\cr},
\label{matrix}
\end{equation}
where $a_{n}=x_{n+1}-x_{n}$ and $\beta $ is defined by Eq. (\ref{polarizability}). The eigen quasi-states for a finite system, $n\in (1,N)$, can be found if one looks for non-trivial solutions when no  wave is incident upon the system. This corresponds to the boundary conditions of the form 
\[
v_{0}=\pmatrix{r\cr -ir\cr} \;  {\rm and} \;\;   v_{N}=\pmatrix{t\cr it\cr}.
\]
A resulting dispersion equation for the eigenfrequencies can be expressed in terms of the elements of the total transfer matrix $\widehat{T}$:
\begin{equation}
\left( T_{11}+T_{22}\right) -i\left( T_{12}-T_{21}\right) =0.
\label{quasi_mode}
\end{equation}
It is clear that for finite samples the solutions of this equation are complex, reflecting the fact that the modes are not stationary; they have lifetimes which tend to infinity only in the limit $N\rightarrow \infty$. 


\begin{figure}
\centering
\vspace{-0.0in}
\epsfxsize=3in \epsfbox{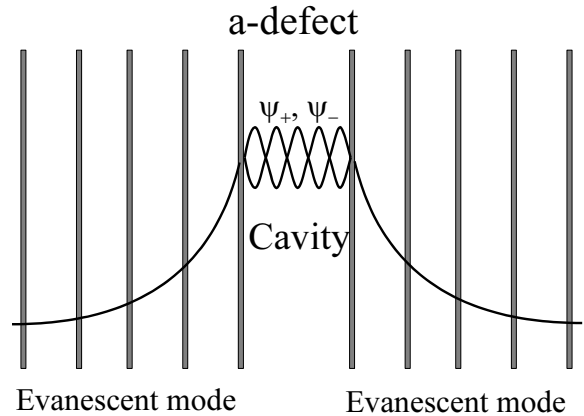}
\vspace{0.2in}
\caption{Matching the solutions for half-infinite perfect MQW with cavity modes $\psi _{\pm }$ one can obtain the dispersion equation for (quasi)local modes (in finite systems) of such a system with the defect.}
\end{figure}

Eq. (\ref{quasi_mode}) for the $a$-defect in the infinite MQW system, after some cumbersome but straightforward algebra, can be presented in the form
\begin{equation}
\cot(k_0 b)=-\frac{\sin (k_0 a)-\beta\lambda _-/2}{ cos (k_0 a)-\lambda _{-}},
\label{cond_inf_a}
\end{equation}
where $\lambda _{-}=\left[ \cot (k_0 a)+\beta/2 - \sqrt{D}\right] \sin (k_0 a)$ is one of the eigenvalues of the transfer matrix, Eq. (\ref{matrix}), and $D=-1+\beta ^{2}/4+\beta \cot (k_0 a)$. This equation, as well as in the cases of $\Gamma $- and $\Omega $-defects, has two solutions - above and below $\Omega_0$. These solutions can be approximated as
\begin{equation}
\omega _{def}^{(1)}=\Omega _{0}-\frac{\omega _u-\omega
_l}{2}\frac{(-1)^{ \left[ \frac{\xi +1}{2} \right] } \sin(\pi\xi/2)
}{1+ \displaystyle{\frac{\omega _{u}-\omega _{l}}{2\Omega _{0}}}\xi
(-1)^{ \left[ \frac{\xi +1}{2} \right] } \cos (\pi\xi/2)   },
\label{om1_a_def}
\end{equation}
\begin{equation}
0=\omega _{def}^{(2)}=\Omega _{0}+\frac{\omega _{u}-\omega
_{l}}{2}\frac{(-1)^{ \left[ \frac{\xi +1}{2} \right] } \cos
(\pi\xi/2)}{1- \displaystyle{\frac{\omega _{u}-\omega _{l}}{2\Omega
_{0}}}\xi  (-1)^{ \left[ \frac{\xi +1}{2} \right] } \sin(\pi\xi/2)  },
\label{om2_a_def}
\end{equation}
where $\xi =b/a$, and $[...]$ denotes an integer part. Therefore, for $\Gamma_0 \ll \Omega_0$ and not very large $\xi$, $\xi \ll (\Omega _0 /\Gamma _0)^{1/2}\simeq 10^2$, and both solutions are almost periodic functions of $b/a$ with the period of $1$, as shown in Fig. 2b.
These solutions oscillate between respective boundaries of the gap ($\omega_u$ or $\omega_l$) and the exciton frequency $\Omega_0$. At the integer values of $\xi$, one of the frequencies $\omega_{def}^{(1)}$ or $\omega_{def}^{(2)}$ becomes equal to $\Omega_0$,  and the other reaches $\omega_u$ or $\omega_l$ depending upon the parity of $\xi$. When $\xi$ crosses an integer value, the solution passing through $\Omega_0$ changes continously, while the second one experiences a jump toward the opposite gap boundary. The observable manifestations of the defect modes (for instance, transmission resonances as described in the next section of this paper) vanish when the defect frequencies approach the gap  boundaries. This jump, therefore, would manifest itself as a disappearance of the transmission peak near one of the gap boundaries, and a gradual reappearance at the opposite edge as $\xi$ changes through an integer value. It is interesting to note again that the exciton resonance frequency $\Omega_0$, which formally lies at the center of the gap, behaves as one of the gap boundaries. This is another manifestation of the fact that the polariton gap in the Bragg structures is formed by two adjacent gaps with a degenerate boundary at $\Omega_0$. 

Calculations of the defect frequencies for the $b$-defect can be done in the framework of both schemes. The transfer-matrix approach, however, turns out to be less cumbersome. The dispersion equation for LPM's in this case takes the form
\begin{eqnarray}
&&\frac{\beta ^2 }{2}\sin ^{2}[k_0(b-a)]\left[ -1+ \frac{\beta }{2\sqrt{D}}+\frac{\cot (k_0 a) }{2\sqrt{D}}\right] \nonumber \\
&+& \cos ^2(k_0 a)+ \sin ^2(k_0 a)\left( \frac{\beta }{2}-\sqrt{D}\right)^2 + \sin (2k_0 a)\left( \frac{\beta }{2}-\sqrt{D}\right)=0. \label{om_b_def_}
\end{eqnarray}
This equation can be solved approximately assuming that the splitting of the solution from the Bragg frequency, $\Omega_0$, is much smaller than the width of the gap. Expanding different terms of Eq. (\ref{om_b_def_}) in terms of powers of $(\omega-\Omega_0)$ and keeping the lowest non-zero contribution, one obtains two different solutions for frequencies of LPM's:
\begin{equation}
\omega _{def}^{(1,2)}=\Omega _{0}
\left[ 1\pm \left( \frac{2}{\pi}\right)^{1/4} \left( \frac{\Gamma_0}{\Omega_0 }\right)^{3/4}\left| \sin {\pi (\xi -1)}\right| \right].
\label{om_b_def}
\end{equation}
The relative splitting of these modes from $\Omega _0$ is of the order of $( \Gamma_0/\Omega_0 )^{3/4}$, and is much smaller than   the relative width of the gap, which is proportional to $( \Gamma_0/\Omega_0 )^{1/2}$ in accord with our initial assumption. Similar to the case of the $a$-defect, these frequencies change periodically with $\xi$, but unlike the previous case, they both split off the center of the gap, $\Omega_0$, and at integer values of $\xi$ merge back to $\Omega_0$. The maximum deviations of the local frequencies from $\Omega_0$ are of the order of $(\Gamma /\Omega_0)^{3/4}$, they take place for half-integer values of $\xi$.

One can notice that the $b$-defect involves two adjacent wells, and in the Green's function approach one would have to deal with a system of two coupled equations. Accordingly, it can be expected that there must be four possible local frequencies (two in each half of the gap), while we found only two of them. The reason for this is that the Bragg condition makes a pair of frequencies, which are both above or below $\Omega_0$, nearly degenerate, and the difference between them is much smaller than the terms we kept in our approximate solution. This can be understood if one notices that the transfer matrices in Eq. (\ref{matrix}) contain  factors $\cos (kb)$, $\sin(kb)$ and $\cos [k(2a-b)]$, $\sin[k(2a-b)]$ for the first and the second of the involved wells, respectively. Exactly at the Bragg frequency, $ka=\pi$, these factors coincide leading to the degeneracy. Since the shift of the  actual local frequency from the Bragg frequency is relatively small, the solutions remain nearly degenerate.  The fact that obtained solutions are symmetrical with respect to the replacement of $b$ with $2a-b$ is not surprising and reflects the symmetry of the transfer matrices discussed above.

\section{Defect LPM's and transmittance and reflectance experiments}

In this Section we study how the local defect modes obtained above affect the reflection and transmission spectra of the finite size periodic Bragg MQW's in the presence of homogeneous broadening.
Transmission and reflection coefficients can be expressed in terms of the elements of the total transfer matrix defined by Eq. (\ref{transfer}) as
\begin{equation}
T=\left| t\right| ^{2}=\left| \frac{2\det \widehat{T}}{\left( T_{11}+T_{22}\right) -i\left(T_{12}-T_{21}\right) }\right| ^{2},
\label{tr}
\end{equation}
\begin{equation}
R=\left| r\right| ^{2}=\left| -\frac{\left( T_{11}-T_{22}\right) +i\left( T_{12}+T_{21}\right) }{\left( T_{11}+T_{22}\right) -i\left( T_{12}-T_{21}\right) }\right| ^{2}.
\label{rf}
\end{equation}
Without absorption, $T$ and $R$ in the form of Eqs. (\ref{tr}) and (\ref{rf}) can be shown to add up to unity. In the denominators of $T$ and $R$ one can recognize the dispersion relation Eq. (\ref {quasi_mode}) that was obtained in the previous section by matching the decaying solutions on both sides of the defect. 

$\Gamma$- and $\Omega$-defects differ from the original QW's only in the definition of the parameter $\beta $, therefore, they both can be dealt with at the same time.  The exact expression for $T$ is cumbersome, however, if  the length of the  system is longer than the penetration length over the frequency region of interest, it can be simplified. In this case the smaller eigenvalue of the total transfer matrix is proportional to $\exp (-Na/l_{loc}(\omega))$, where $l_{loc}$ is given by Eq. (\ref{extinction_length}), and can be neglected. The expression for the transmission coefficient then can be presented as\cite{Deych}
\begin{equation}
t=\frac{t_{0}}{\left( 1+\varepsilon \right) +i\exp \left( -ik_0 Na\right) \varepsilon \Phi t_{0}\cosh \left[ (N-2n_{0}+1)\kappa a\right] }, 
\label{trans1}
\end{equation}
where we introduce the defect parameter $\varepsilon =\left( \beta _{def}-\beta \right) /2\sqrt{D}$, which is equal to zero when $\beta _{def}=\beta $; $n_{0}$ is the position of the defect QW, $\kappa =1/l_{loc}(\omega )$, and $\Phi =\beta /(\sin (k_0 a)\sqrt{D})$. For $\varepsilon =0,$ Eq. (\ref{trans1}) gives the transmission coefficient, $t_{0},$ of the pure system, 
\begin{equation}
t_{0}=\frac{2e^{ikL}\exp \left( -\kappa L\right) }{1+i\left[ 2-\beta \cot( ka)\right]/\sqrt{D} },  
\label{t_pure}
\end{equation}
exhibiting an exponential decay characteristic of the evanescent modes from a band gap.

Eq. (\ref {trans1}) describes the resonance tunneling of the electromagnetic waves through MQW with the defect. The equation
$$
1+\varepsilon=0
$$
can be shown to coincide with the dispersion equations for local defect modes of the infinite structure in the case of $\Omega$ and $\Gamma$ defects. Given this fact and the structure of Eq. (\ref{trans1}), it seems natural to assume, as we did in our previous papers,\cite{Deych} that the transmission reaches its maximum value at the frequencies of the local modes. More careful consideration showed, however, that the systems under consideration behave in a less trivial way, and that maxima of the transmission occur at frequencies shifted from the frequency of the local modes. This is quite unusual behavior that distinguishes the systems under consideration from other instances of resonance tunneling. Moreover, we shall show that contrary to our previous result,\cite{Deych} the transmission at the maximum is always equal to unity (in the absence of absorption). 

Evaluation of the exact analytical expression for the transmission coefficient, Eq. (\ref{tr}), obtained from the total transfer matrix, showed that the distinctive resonance transmission occurs only if local modes lie not too close to the boundaries or the center (which is, strictly speaking, also a degenerate boundary) of the gap. $\Gamma$ defects, therefore, can be excluded from consideration, as well as one of the solutions for the $\Omega$-defect. For frequencies close to the remaining local mode (the solution for the $\Omega$-defect given by Eq. (\ref{cond_inf_Om_app}) the expression for the transmission coefficient Eq. (\ref{trans1}) can be simplified.  Expansion in the vicinity of the local mode gives
\begin{equation}
T=\frac{\displaystyle{\left(\frac{\omega-\Omega_1}{\omega_{def}-\Omega_1}\right)^2}}{1+A_1 \left|\displaystyle{ \frac{\omega -\omega_T\cosh \left( (N-2n_{0}+1)\kappa a\right)(1+i\cdot A_2 )}{\omega_{def}-\omega_T}}\right|^2 \cdot \displaystyle{\left(\frac{\omega_T-\Omega_1}{\omega_{def}-\Omega_1}\right)^2}} \cdot  
\label{transm2}
\end{equation}
Parameters $A_1$, $A_2$ and $\omega_T$ are defined by
\begin{equation}
A_1=\frac{(\omega_{+} -\omega_{def})^2(\omega_{def} -\omega_{-})^2}{4(\omega_{def} -\Omega_0)^2(\omega_{def} -\omega_l)(\omega_u -\omega_{def})},\label{A1}
\end{equation}
\begin{equation}
A_2=\frac{(\omega_{def} -\Omega_0)\sqrt{(\omega_{def} -\omega_l)(\omega_u -\omega_{def})}}{(\omega_{+} -\omega_{def})(\omega_{def} -\omega_{-})} \tanh [ (N-2n_{0}+1)\kappa a ] ,\label{A2}
\end{equation}
\begin{equation}
\omega_T=\omega_{def}+\pi\frac{(\omega_{+} -\omega_{def})(\omega_{def} -\omega_{-})}{\Omega_0}\frac{(\omega_{def} -\Omega_0)}{\sqrt{(\omega_{def} -\omega_l)(\omega_u -\omega_{def})}}e^{-\kappa Na} ,
\label{omegaT}
\end{equation}
where $\omega _{\pm}=\Omega _{0}\pm \left( \omega _{u}-\Omega _{0}\right)/\sqrt{2} $. The resonance transmission occurs when the defect layer is located in the center of the system $N-2n_0+1=0$. In this case the coefficient $A_2$ becomes zero and Eq. (\ref{transm2}) can be presented in the following form
\begin{equation}
T=\frac{4\gamma_{\Omega} ^2}{Q^2}\frac{(\omega -\omega_T+Q)^2}{(\omega -\omega_T)^2+4\gamma^2_{\Omega }} ,  
\label{transm3}
\end{equation}
where $Q=\omega_T-\Omega_1$, and the parameter $\gamma_{\Omega} $ is given by
\begin{equation}
\gamma_{\Omega}=\pi \Omega _0 \left( \frac{\omega_{def} -\Omega_0}{\Omega_0 }\right) ^2 e^{-\kappa Na}.  
\label{Ogamma}
\end{equation}
The transmission spectrum described by Eq. (\ref{transm3}) has a shape known as a Fano resonance,\cite{Fano} where $\omega_T$ is the resonance frequency, at which the transmission turns to unity, and parameters $\gamma_{\Omega}$ and $Q$ describe the width and the asymmetry of the resonance respectively. One can see from Eq. (\ref{omegaT}) that in general the transmission resonance frequency is shifted with respect to the frequency of the local mode. The shift, though exponentially small for long systems considered here, is of the same order of magnitude as the width of the resonance $\gamma_{\Omega}$, and is, therefore, significant. These two frequencies, $\omega_T$ and $\omega_{def}$, coincide in the special case when $\omega_{def}=\omega_{\pm}$. The fact that the transmission is equal to one, in this particular case, was obtained in our previous paper Ref. \onlinecite{Deych}. 

\begin{figure}
\centering
\vspace{-0.0in}
\epsfxsize=3in \epsfbox{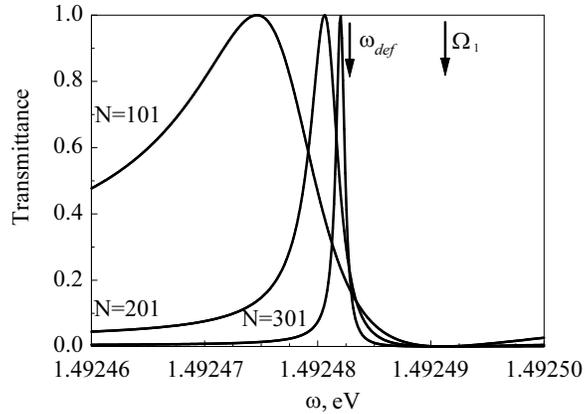}
\vspace{0.2in}
\caption{The shape of the transmission maximum for 3 lengths $N=$ 101, 201, 301. For all lengths, the $\Omega $-defect is in the center of the system.  $\Omega_1=1.492491$ eV is the exciton frequency of the defect layer. The exciton resonant frequency $\Omega_0$ is $1.491$ eV, the exciton-light coupling constant $\Gamma _0$ is $27\;\mu$eV as in the InGaAs/GaAs structures studied in Ref. 8, however, no homogeneous broadening is assumed.}
\end{figure}

At  $\omega=\omega_T-Q=\Omega_1$ the transmission equals zero, which is a signature of Fano resonances. Usually the presence of Fano resonances is associated with interaction between a discrete level and a continuum of states. This is not the case  in the situation under consideration.  Zero of the transmission in our case is caused by the fact that the $\Omega$-defect brings to the dielectric function of the system a new  pole at $\Omega_1$, which describes the interaction between electromagnetic waves and excitons  of the defect well. The penetration length diverges at $\Omega_1$, and transmission, therefore, vanishes.  The interaction with excitons, at the same time, leads to the radiative shift of the frequency of the local mode, and hence, the frequency of the transmission resonance, away from $\Omega_1$. The combination of these two factors is responsible for the Fano-like shape of the transmission resonance.  The actual form of the Fano spectrum in an ideal system without absorption is determined by the interplay between the width parameter $\gamma_{\Omega}$ and the asymmetry parameter $Q$. The former exponentially decreases with an increase of the length of the system, while the latter is length independent. However, the pre-exponential factor in $\gamma_{\Omega}$ is of the order of the exciton resonance frequency, $\Omega_0$, while $Q$ is of the order of $\Gamma$, i.e. significantly smaller. Therefore, in principle, there are two possible cases: $\gamma_{\Omega}\gg Q$ for shorter systems [$1\ll \kappa Na\ll \log(\Omega_0/Q)$], and $\gamma_{\Omega}\ll Q$ for longer systems.  In the first case, the transmission spectrum has a distinctively Fano-like asymmetrical shape, while in the second limit the spectrum attains the symmetrical Lorentzian shape characteristic for Wigner-Breight\cite{WB} resonances. Fig. 4 shows the evolution of the shape of the transmission resonance in the absence of absorption from Fano-like to Wigner-Breight-like behavior with increase of the length of the system. One can also see from this figure how the position of the transmission maximum moves with an increase of the length. An actual possibility to observe the Fano resonance in the considered situation depends strongly upon the strength of absorption in the system, which must be at least smaller than $Q$. More detailed discussion of absorption related effects is given below in the present section.

Calculation of the transmission coefficient for the $a$-defect can be carried out in a similar manner. Dropping the exponentially small contribution from the smaller eigenvalue the transmission matrix can be written in the following form:
\begin{equation}
t=\frac{t_{0}}{\left( 1+\varepsilon _{a} \right) +ie^{-i\frac {\omega }{c}Na}\Phi ^2\sin (\frac {\omega }{c}(b-a)) t_{0}\cosh \left[ (N-2n_{0})\kappa a\right] },  
\label{trans2}
\end{equation}
where 
\[ \varepsilon _{a}=\frac {\sin (\frac {\omega }{c}b)-\sin (\frac {\omega }{c}a)-\lambda _{+}\left( 1-\frac {\beta }{2\sqrt{D}}\right) \sin (\frac {\omega }{c}(b-a))}{\sin (\frac {\omega }{c}a)}. \] 
Similar to the previous defects, $1+\varepsilon _{a} =0$ coincides with the dispersion equation for an infinite system, but the transmission resonance is shifted with respect to the local mode. In the vicinity of the resonance, Eq. (\ref{trans2}) can be presented as: 
\begin{equation}
T=\frac{1}{1+A_1 \left|\displaystyle{ \frac{\omega -\omega_T\cosh \left( (N-2n_{0})\kappa a\right)(1+i\cdot A_2 )}{\omega_{def}-\omega_T}}\right|^2 },  
\label{atransm}
\end{equation}
where parameters $A_1$ and $A_2$ are again given by Eqs. (\ref{A1}) and (\ref{A2}), respectively. One only needs to replace $N-2n_0+1$ in Eq. (\ref{A2}) with $ N-2n_0$, which reflects the new symmetry of the system. The frequencies of local modes  $\omega _{def}$ are now given by Eqs. (\ref{om1_a_def}) and (\ref{om2_a_def}), and $\omega_T$ is defined by
\begin{equation}
\omega_T=\omega_{def}+2(\omega_{def}-\Omega_0)
\frac{(\omega_{+} -\omega_{def})(\omega_{def} -\omega_{-})(\omega_{def} -\omega_l)(\omega_u -\omega_{def})}{(\omega_u-\Omega_0)^4}e^{-\kappa Na}.
\label{aomegaT}
\end{equation}
The resonance transmission again occurs when $ N-2n_0=0$, which requires an even number of wells in the system. Eq. (\ref{atransm}) in this case takes the standard Wigner-Breight shape
\begin{equation}
T=\frac{4\gamma_a ^2}{(\omega -\omega_T)^2+4\gamma_a^2},  
\label{atransm1}
\end{equation}
with the half-width, $\gamma_a $, given now by
\begin{equation}
\gamma_a=2  \frac{(\omega_{def} -\Omega_0)^2(\omega_{def} -\omega_l)^{3/2}(\omega_{u} -\omega_{def})^{3/2}}{(\omega_u-\Omega_0 )^4} e^{-\kappa Na},  
\label{agamma}
\end{equation}
The frequency $\omega_T$, where the transmission coefficient takes the maximum value of unity, is again shifted from the frequency of the defect mode. The two frequencies coincide, however, when the defect frequency is made equal to $\omega_{\pm}$, which are the same frequencies at which transmission resonance and the local mode coincide for the $\Omega $-defect. As one can see from Eq. (\ref{om1_a_def}), these conditions can be satisfied simultaneously for both defect frequencies of the $a$-defect when $b \simeq (integer+1/2)a$ (see Fig. 2b). In this case the position of the transmission resonance becomes independent of the length of the system.

Finally, the $b$-defect gives an expression for the transmission coefficient very similar to Eq. (\ref{trans1}) with the only distinction that $\epsilon $ has to be replaced by a different expression, which is too cumbersome to be displayed here. The maximum transmission for a given defect is again achieved when the defect is in the center of the system (N is odd). Expanding the transmission coefficient near the frequency of the respective local mode, one obtains
\begin{equation}
T=\frac{1}{1+\displaystyle{ \frac{1}{4}\left( \frac{\omega_{u}-\Omega_0}{\omega_{def}-\Omega_0}\right)^2 \left| \frac{\omega -\omega_T}{\omega_{def}-\omega_T}+i\cdot \sinh \left( (N-2n_{0}+1)\kappa a\right)\right|^2} },
\label{trans3}
\end{equation}
with $\omega_T$ now given by
\begin{equation}
\omega_T=\omega_{def}-2  (\omega_{def} -\Omega_0 )e^{-\kappa Na}.
\label{bomegaT}
\end{equation}
Unlike other types of defects, in this case the transmission resonance is always different from $\omega_{def}$. When the defect is at the resonance position $N-2n_{0}+1=0$, the transmission spectrum again takes the Wigner-Breight shape with the resonance width determined by parameter  $\gamma_b$, 
where $\gamma_b$ is given by 
\begin{equation}
\gamma _b=2  \frac{(\omega_{def} -\Omega_0 )^2}{\omega_{u} -\Omega_0 }
e^{-\kappa Na},
\label{bgamma}
\end{equation}

In real systems, enhancement of the transmission coefficient is usually limited
by homogeneous broadening of exciton resonances. Two cases are possible when
exciton damping is taken into account. It can suppress the resonance
transmission, and the presence of the local states can only be observed in not
very long systems as a small enhancement of absorption at the local
frequency. This can be called a weak coupling regime for LPM, when incident
radiation is resonantly absorbed by local exciton states. The opposite case,
when the resonance transmission persists in the presence of damping, can be
called a strong coupling regime. In this case, there is a coherent coupling
between the exciton and the electromagnetic field, so that the local states can
be suitably called local polaritons. Qualitatively we can assess the effect of
absorption on resonances caused by different defects by looking at the widths
of the respective spectra. For all types of defects the width of respective
resonances exponentially decreases with the length of the system, consequently,
in sufficiently long systems all resonances dissappear. However,
pre-exponential factors make different defects behave differently at
intermediate distances. A simple qualitative estimate would require that the
width of the resonances be smaller than the exciton relaxation
parameter. Therefore, the resonances, where the pre-exponential factor of the
width is considerably larger than the relaxation parameter, can be observed in
the systems of intermediate length. On the one hand, the length must be greater
than the localization length of the respective local mode. On the other hand,
it must be small enough for the width of the resonance to remain larger than
the exciton relaxation parameter. The Fano resonance arising in the case of the
$\Omega$-defect though, requires special consideration since its vitality is
determined by the asymmetry parameter $Q$ rather than by the width parameter
$\gamma_{\Omega}$. Although the latter is determined by the large pre-exponent
(of the order of the exciton resonance frequency $\Omega_0$), the former is of
the order of the light-exciton coupling constant $\Gamma_0$, which is much
smaller. The Fano resonance will likely be washed out as soon as the relaxation
rate exceeds this asymmetry parameter $Q$. This circumstance will prevent
observation of the Fano resonance due to $\Omega$-defect in $InGaAs/GaAs$
MQW's, experimentally studied in Ref. \onlinecite{Khitrova} which is, to the
best of our knowledge, the only system, where radiative coupling was observed
for systems as long as 100 wells. In this system, the exciton resonance
frequency, $\Omega_0$, and the exciton-light coupling parameter, $\Gamma_0$,
were respectively equal to $\Omega _{0}=1.491\;eV$, and $\Gamma _{0}=27\;\mu
eV$, while the exciton relaxation parameter was estimated as $\gamma_{hom}
=280\;\mu eV$\cite{Khitrova}.  However, the Bragg $GaAs/AlGaAs$ MQW system
studied in Ref.\onlinecite{Hubner} offers much more favorable conditions for
the observation of this effect. According to Ref.\onlinecite{Hubner} radiative
and non-radiative rates in $GaAs/Al_{0.3}Ga_{0.7}As$ system consisting of 10
wells are equal respectively to $\Gamma _{0}=67\;\mu eV$ and
$\gamma_{hom}=12.6\;\mu ev$. If one could grow sufficiently long (60 - 80 wells
or more) $GaAs/AlGaAs$ MQW's with the similar relation between $\Gamma_0$ and
$\gamma_{hom}$, the observation of the Fano resonance induced by the
$\Omega$-defect may be experimentally feasible.

\begin{figure}
\centering
\vspace{0.0in}
\epsfxsize=2.6in \epsfbox{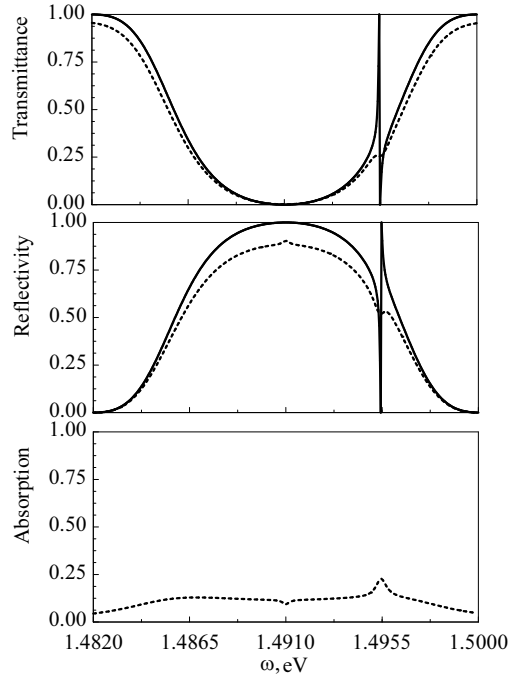}
\vspace{0.2in}
\caption{Transmission, reflection and absorption coefficients for the $\Omega $-type defect. The defect is placed in the center of the MQW with 201 quantum wells. The exciton frequency of the defect well $\Omega_1$ is chosen such that $(\Omega _1 -\Omega_0 )/\Omega_0 =1.003$. Numerical values of the exciton resonant frequency and the exciton-light coupling constant were taken for $InGaAs/GaAs$ structures studied in Ref. 8. Solid lines show results obtained in the absence of absorption, and dashed lines were calculated in the presence of experimentally observed\cite{Khitrova} homogeneous broadening.}
\end{figure}

The pre-exponent of the resonance width in the case of the $a$-defect is of the order of magnitude of the gap width, which is proportional to $\sqrt{\Gamma_0\Omega_0}$. It is considerably larger than $\gamma_{hom}$ for the $InGaAs/GaAs$ MQW's and we expect, therefore, that it can be easily observed in readily available samples of this composition. As far as the $b$-defect is concerned, it gives rise to extremely narrow transmission peaks, which are characterized by a pre-exponential factor of the order of $\Gamma_0$. It, therefore, will be washed out in $InGaAs/GaAs$ MQW's, but can be reproduced in $60 - 80$ wells long $GaAs/AlGaAs$ systems with the parameters similar to those reported in Ref.\onlinecite{Hubner}.  

\begin{figure}
\centering
\vspace{0.0in}
\epsfxsize=2.5in \epsfbox{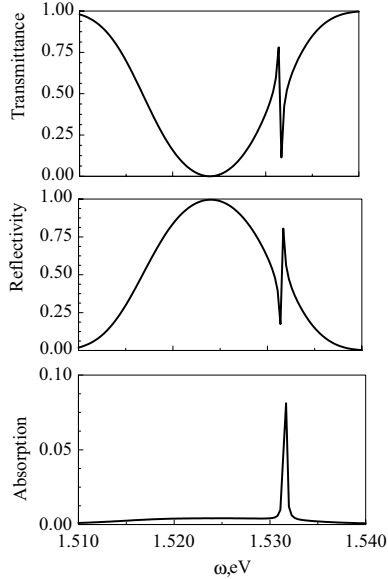}
\vspace{0.0in}
\caption{Transmission, reflection and absorption coefficients for the $\Omega $-type defect. The defect is placed in the center of the MQW with 101 quantum wells. The exciton frequency of the defect well $\Omega_1$ is chosen such that $(\Omega _1 -\Omega_0 )/\Omega_0 =1.003$. Numerical values of the exciton resonant frequency and the exciton-light coupling constant were taken for $GaAs/AlGaAs$ structures studied in Ref. 5. All results were obtained in the presence of experimentally observed\cite{Hubner} homogeneous broadening.}
\end{figure}

We compliment the qualitative arguments presented above with numerical evaluation of exact expressions for the transmission, and reflection coefficients, Eqs. (\ref{tr}) and (\ref{rf}), using parameters of the mentioned $InGaAs/GaAs$ and $GaAs/AlGaAs$ systems. To account for homogeneous broadening quantitatively, we add an imaginary part to the denominator of the parameter $\beta$:
\[
\beta =\frac{4\Gamma _{0}\omega}{\omega ^{2}-\Omega _{0}^{2}+2i\gamma_{hom} \omega}.
\]
Fig. 5 shows transmission and reflection coefficients of the Bragg MQW lattice made of 201  quantum wells with the $\Omega $-defect in the center with parameters corresponding to the $InGaAs/GaAs$ system. One can see that absorption washes out the strong asymmetric pattern of the Fano resonance, but some remains of the resonance are still quite prominent, and can probably be observed in high quality samples. Fig. 6 represents the spectra of 101 well long system with parameters corresponding to  $GaAs/AlGaAs$. One can see a characteristic assymetric profile of the Fano resonance both in transmission and reflection. There is also a remarkably strong and narrow absorption line at the resonance frequency with more than 8-fold growth of absorption at the resonance. This behavior should be contrasted with the previous figure, where absorption spectrum is rather flat with only insignificant increase at the resonance frequency. We confirm, therefore, our conclusion that  $GaAs/AlGaAs$ systems with the $\Omega$-defect might have a potential for realization of local polariton modes in the regime of strong coupling. In the case of the $b$-defect, strong exciton absorption characteristic for the $InGaAs/GaAs$ structures washes out any resonance features from the spectrum. The resonances, however, survive for a system with parameters similar to those reported in Ref.\onlinecite{Hubner} for $GaAs/AlGaAs$. In this case one has very narrow symmetric transmission, reflection and absorption lines (Fig.7) located rather close to the center of the gap.  It is interesting to compare resonance behavior of the $b$-defect and the $\Omega$-defect in the regime of strong coupling. The latter results in  absorption that, though increased sharply at the resonance, is still rather weak, at the same time the transmission reaches in this case almost $80\%$. In the former case, however, the transmission does not grow that dramatically (up to about $0.4$), but absorption increases by two orders of magnitude. The difference is caused by the different shapes of the resonances: the Fano resonance is considerably wider and, therefore, results in less dramatic increase in the maximum absorption. 

\begin{figure}
\centering
\vspace{-0.1in}
\epsfxsize=2.5in \epsfbox{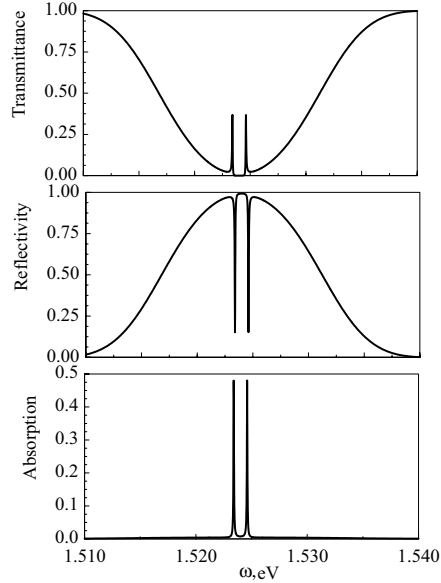}
\vspace{-0.1in}
\caption{Transmission, reflection and absorption coefficients for the $b$-type defect. The defect is placed in the center of the MQW with 101 quantum wells. The defect strength is $\xi =b/a=1.5$. Numerical values of the exciton resonant frequency and the exciton-light coupling constant were taken for $GaAs/AlGaAs$ structures studied in Ref. 5. All results were obtained in the presence of experimentally observed\cite{Hubner} homogeneous broadening.}
\end{figure}

Fig. 8 shows the transmission and reflection of the Bragg MQW lattice of 200 quantum wells with the $a$-defect with $\xi=b/a=1.5$ and parameters of the $InGaAs/GaAs$ system. This is the only defect for which the strong coupling regime can be realized for  this material. But since this is the only system, for which the samples with large number of wells were experimentally grown, we shall discuss the related results in more details. 

\begin{figure}
\centering
\vspace{0.0in}
\epsfxsize=2.5in \epsfbox{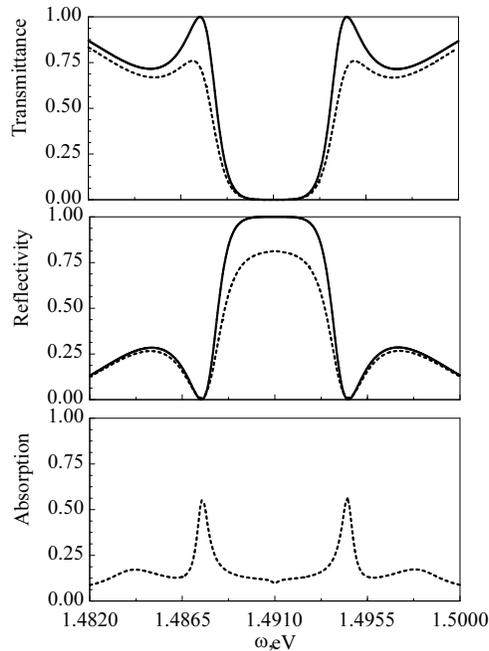}
\vspace{0.1in}
\caption{Transmission, reflection and absorption coefficients for the $a$-type defect. The defect is placed in the center of the MQW with 200 quantum wells. The defect strength is $\xi =b/a=1.5$. Numerical values of the exciton resonant frequency and the exciton-light coupling constant were taken for $InGaAs/GaAs$ structures studied in Ref. 8. Solid lines show results obtained in the absence of absorption, and dashed lines were calculated in the presence of experimentally observed\cite{Khitrova} homogeneous broadening.}
\end{figure}

For the $a$-defect with the strength $\xi=1.5$, one has two symmetric with respect to the center resonances. The peaks are very pronounced in the tranmission, and the contrast in reflection is very large - about 30\%, and can be made even larger if one increases the length of the system. The spectra shown in Fig. 8  can, however, be affected by the presence of the additional resonance due to $2s$ excitons.  In Ref.\onlinecite{Khitrova} it was shown that this resonance causes asymmetry of the reflection spectrum. We expect that it could have the similar effect upon spectra presented in Fig. 8 causing an assymetry between low- and high- frequency portions of the spectra. Even if $2s$ excitons  contribute additional resonances to the spectra, they would be rather weak,\cite{Khitrova} and could be clearly distinguished from the effects considered in the present work. As an additional identification tool, one can use the dependence of the defect induced features upon the defect parameter $b/a$, while a contribution from the $2s$ exciton resonance would not be affected by changes in this parameter. Besides, changing the parameters of the defect, one can move the transmission resonances closer to the center of the gap and farther away from the $2s$ exciton resonance.

We can conclude, therefore, that the strong coupling between local excitons and local photons can be experimentally realized in readily available samples of $InGaAs/GaAs$ MQW's. If, however, one could grow long systems of $GaAs/AlGaAs$ MQW's with parameters close to those reported in Ref.\onlinecite{Hubner}, the defect induced optical resonances would be even more pronounced with more dramatic increase in transmission and absorption at the resonance frequencies.   

\begin{figure}
\centering
\vspace{0.0in}
\epsfxsize=2.5in \epsfbox{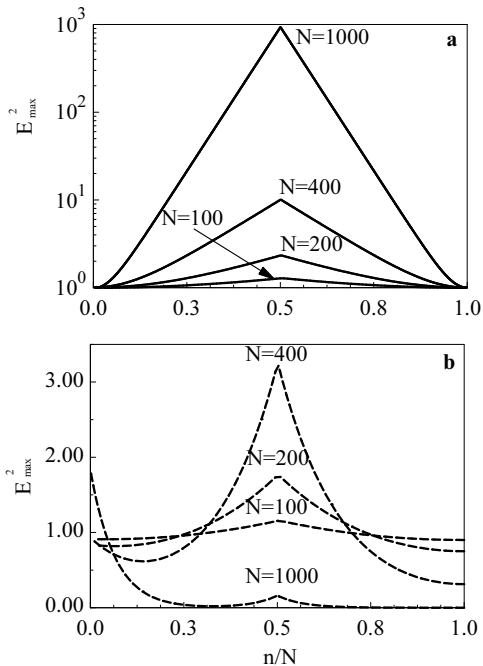}
\vspace{0.1in}
\caption{The distribution of the electric field at the frequency $\omega_T$ in the system with the $a$-defect placed exactly in the middle without (a) and with (b) experimentally observed\cite{Khitrova} homogenous broadening. The defect strength is $\xi =b/a=1.5$. Numerical values of the exciton resonant frequency and the exciton-light coupling constant were taken for $InGaAs/GaAs$ structures studied in Ref. 8. Four curves correspond to different sizes of the system: 100, 200, 400 and 1000 QW's.}
\end{figure}

It is interesting to note that the effects of local modes on the resulting absorption of the incident wave is very weak in the case of weak coupling. Even though there is some small enhancement of absorption at the resonance frequencies, it is much weaker than absorption peaks when strong coupling is realized. In other words, local polaritons in the regime of strong coupling demonstrate resonance behavior in both transmission and absorption at the same time, while in the weak coupling regime there is only a small effect of the local states upon all optical spectra of the system. The explanation for this behavior lies in the spatial distribution of the electromagnetic wave intensity throughout the system. In the absence of absorption, the electric field at the resonance  frequency decays exponentially away from the defect layer, i.e. there occurs a strong exponential enhancement of the incident field at the defect layer.\cite{PRBlocal,Azbel} We used self-embedding technique, adopted for the discrete systems in Ref.\onlinecite{PRBlocal}, to calculate numerically the electromagnetic field inside MQW structure. In the absence of absorption we observed the exponential increase of the field compared to the amplitude of the incident electromagnetic wave $E_{in}$
\begin{equation}
|E_{max}|=|E_{in}| \cdot e^{(N/2)\kappa a}.
\label{Emax}
\end{equation}
Non-radiative broadening suppresses not only the resonance transmission but also the exponential enhancement of the electric field. If the strong coupling regime is not realized, the intensity of the wave decreases exponentially throughout the sample almost as it would in the absence of the defect, and is just slightly larger at the resonance frequency  than off-resonance. Therefore, the peaks in absorption in these cases are also only minute. At the same time, Fig. 9 shows that even in the presence of absorption the $a$-defect in the $InGaAs/GaAs$ MQW's, which remains in the regime of strong coupling, demonstrates more than a three-fold enhancement of $|E|^2$ at the location of the defect for the system of an optimal length. Fig. 10a shows the evolution of the electric field at frequency  $\omega _T$ at the location of the defect as the size of the system grows (for systems with different sizes the defect is always located in the center of the structure). One can see that the exponential growth of the field in the absence of absorption (solid curve) changes to a nonmonotonic behavior when absorption is included (dashed line). For the particular system under consideration, the field  reaches its maximum at about $N_m$=450, where we see the crossover from the resonant enhancement regime ($N<N_m$) to the exponential decay ($N>N_m$). Enhancement of the field at the defect explains both transmission and absorption resonances. 

\begin{figure}
\centering
\vspace{0.0in}
\epsfxsize=2.5in \epsfbox{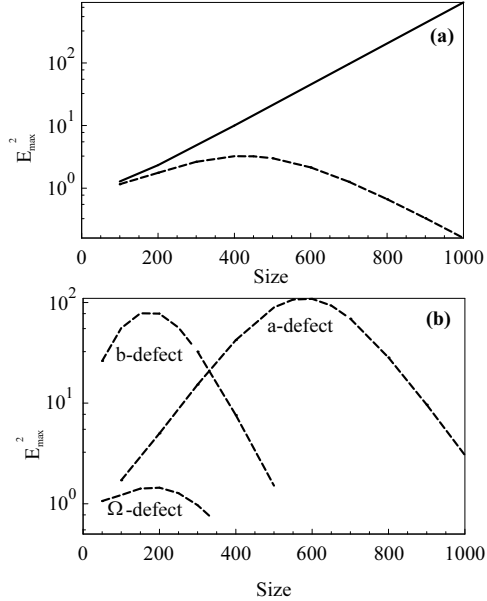}
\vspace{0.1in}
\caption{(a) The dependence of the strength of the electric field at the position of the $a$-defect in $InGaAs/GaAs $ structures (from Fig. 9) at $\omega_T$ on the size of the system with (dashed line) and without (solid line) homogeneous broadening. The other defects did not demonstrate resonance enhancement of the electric field and are not shown here. (b) The dependence of the strength of the electric field at the position of the $a$-, $b$-, and $\Omega$-defects in $GaAs/AlGaAs$ structures at $\omega_T$ on the size of the system is shown. Numerical values of the exciton resonant frequency and the exciton-light coupling constant were taken from Ref. 5. All results were obtained in the presence of experimentally observed\cite{Hubner} homogeneous broadening. The defect strengths are $\xi =b/a=1.5$ for $a$- and $b$-defects, and $(\Omega _1 -\Omega_0 )/\Omega_0 =0.9963$ for the $\Omega$-defect. In the presence of the realistic homogeneous broadening, all 3 types of the defects show the resonance enhancement of the electric field.}
\end{figure}

In the case of $GaAs/AlGaAs$ parameters, one can see from Fig. 10b much stronger enhancement of the field at the resonance for three types of defects. Judging by the strength of the absorption line shown in Fig. 7, one could expect an increase in the field at the $b$-defect by about two orders of magnitude. From Fig. 10b we see that for $N=100$ the intensity of the electric field is, indeed, enhanced by the factor of 50. The difference in absorption between $b$- and $\Omega$-defects also signals about the difference in the field distribution. The fact that the $\Omega$-defect leads to the vanishing transmission at the frequency $\Omega_1$ translates, as computer simulations show, into the vanishing electric field everywhere behind the defect at this frequency -- the defect acts as an almost perfect reflector. The growth of the intensity of the field at spectrally close frequency $\Omega_0$ is also limited to only 1.5 of the incident intensity, which is much smaller than in the case of the $b$-defect. 
The strong enhancement of the field at the location of the $b$-defect has simple physical explanation. The frequency of the local states due to the $b$-defect lies close to the exciton frequency that is tuned such that $\lambda /2=a$. This means that at the frequency of the local mode all QWs but one (the defect one, see also Fig. 1b) lie in the nodes of the electromagnetic wave\cite{Ivchenko}. Therefore, one could expect  stronger then usual grouth of the electric field at the defect. From this considerations,  one should expect the maximum effect to occur when the defect QW is placed in an antinode, where the electric field is the largest. We already encountered this condition, when we studied the transmittance in the presence of this type of defect -- the most favorable value of the defect strength was $b/a=0.5$ or $b/a=1.5$. This peculiar spatial distribution of the electric field explains the narrow transmission peak and strong enhancement of the electric field on the defect. This fact has great potential for applications if this type of the defect is realized experimentally.

\section{Conclusion}

The main objective of the present paper was to demonstrate that introducing ``defects" in periodic Bragg MQW systems, one can obtain new opportunities to tailor optical properties of quantum heterostructures. By ``defects" we mean wells or barriers with properties different from those of the host structure. We consider four different types of defects  and show that each of them affects the optical spectra of the structure differently. Using available from literature parameters of $GaAs/AlGaAs$\cite{Hubner} and $InGaAs/GaAs$\cite{Khitrova} MQW's we generate the reflection and transmission spectra and show that they are significantly modified in the presence of the defect layers. 

Among four types of defect considered in the paper, three are of special interest. First of them, the $\Omega$-defect, arises when  one of the QW is replaced by a well with a different exciton resonance frequency $\Omega_1$. In this case two new local modes arise within the band-gap of the original structure, but only one of them, which manifests radiative shift  from $\Omega_1$, can significantly effect optical spectra of the system. We found that the resonance transmission of radiation due to the local state results in strongly assymetric Fano-like transmission spectrum. This transmission resonance turned out to be very sensitive to the presence of non-radiative broadening, it can only survive if the total broadening (including both homogeneous and non-homogeneous contributions) does not exceed the strength of radiative coupling between excitons and light. Using available data we concluded that the Fano resonance could not be observed in  $InGaAs/GaAs$ systems, but would survive in  $GaAs/AlGaAs$ MQW's provided that one could grow $60-100$ wells long systems with the parameters similar to those reported in Ref.\onlinecite{Hubner} for $10$ wells long structures.  MQW's with these parameters would demonstrate strong assymetric transmission/reflection lines with possible increase in transmission up to 80\%. 
Absorption spectrum of such structure is also characterized by a strong a narrow line with absorption increased at resonance, but is still remanins relatively weak. The weakness of the absoprtion correlates with the fact that the intensity of the field at this defect increases only slightly.  

Another defect with interesting properties considered in the paper is the $b$-defect, which arises when one shifts the position of one of the wells keeping coordinates of all others intact. This corresponds to changes in the thicknesses of two adjacent barriers. This defect produces two local modes with frequencies close to the center of the gap. Respective transmission resonances have regular symmetric Lorentzian shapes but with rather narrow widths. The last circumstance makes these defects also very vulnerable to the non-radiative broadening, and they can only survive in systems with the radiative decay rate greater than the non-radiative one ($GaAs/AlGaAs$ may be an example of such systems\cite{Hubner}). If, however, this resonance is realized, one can obtain a narrow absorption line with almost two orders of magnitude increase in absorption at the resonance, or what is even more important, 50-fold increase in the intensity of the electromagnetic field at the defect QW.

The most robust of the local polariton modes is produced by the $a$-defect. In this case the width of one of the barriers is increased, while all other barriers remain the same. The defect produced in this way resembles a regular microcavity, but with optically active mirrors. This fact distinguishes the system considered here from regular cavities with, for instance, Bragg distributed mirrors. Local states arising in this case give rise to rather wide resonances in optical spectra, which can survive strong enough non-radiative broadening and can be observed in readily available $InGaAs/GaAs$  samples. 

The results obtained demonstrate that one has a great variety of opportunities to tailor optical properties using defect MQW's. These systems can have a range of different applications. For example, the great sensitivity of the defect induced features of optical spectra to characteristics of the system can be used for characterization of both host QW's and defect wells. This sensitivity also leads to possibilities to tune the frequencies of local modes by means of application of  stress,  electric or magnetic field, or other stimuli. Another important feature with great potential for applications is the increase of the field intensity in the vicinity of the defect well. It can be utilized in order to  enhance optical non-linearity of the system. It is also worth noting that the local modes considered here can be useful as tunable sources of narrow luminescent lines at the frequencies of the local modes. 

Besides opportunities for applications, the effects considered in the paper are interesting from the academic point of view.  The tunneling resonances associated with the local polaritons have a number of peculiarities compared to other examples of  resonance tunneling phenomena. For instance, the Fano-like transmission produced by the $\Omega$-defect appears here under new circumstances, specific for this particular system. Another interesting feature of the local polaritons is that the frequencies of transmission resonances are always shifted with respect to the eigen frequencies of the modes and depends upon the length of the system. 

Concluding, the structures considered in the present paper demonstrate a number of interesting optical effects and have a potential for a variety of applications. We hope that this paper would stimulate experimental observation and utilization of the predicted here effects.

\section*{Acknowledgments}

We would like to thank David Citrin and Gottfried Strasser for useful discussions of our work. We are also indebted to S. Schwarz for reading and commenting on the manuscript. This work was partially supported by NATO Linkage Grant N974573, CUNY Collaborative Grant, and PSC-CUNY Research Award.


\begin{references}

\bibitem{Citrin} D. S. Citrin, Solid State Commun. {\bf 89}, 139 (1994).

\bibitem{Ivchenko} E. L. Ivchenko, A. I. Nesvizhskii, and S. Jorda,  Phys. Solid Sate {\bf 36}, 1156 (1994).

\bibitem{Andreani} L. C. Andreani, Phys. Lett. A {\bf 192}, 99 (1994); Phys. Status Solidi B {\bf 188}, 29 (1995).

\bibitem{Bjork} G. Bj\"{o}rk, S. Pau, J.M. Jakobson, H. Cao, and Y. Yamamoto, Phys. Rev. B {\bf 52}, 17310 (1995).

\bibitem{Hubner} M. H\"{u}bner, J. Kuhl, T. Stroucken, A. Knorr, S. W. Koch, R. Hey, K. Ploog,  Phys. Rev. Lett. {\bf 76}, 4199 (1996).

\bibitem{Stroucken} T. Stroucken, A. Knorr, P. Thomas, and S. W. Koch,  Phys. Rev. B {\bf 53}, 2026 (1996).

\bibitem{Vladimirova} M. P. Vladimirova, E. L. Ivchenko, and A. V. Kavokin, Semiconductors {\bf 32}, 90 (1998).

\bibitem{Khitrova} M. H\"{u}bner, J. P. Prineas, C. Ell, P. Brick, E. S. Lee, G. Khitrova, H. M. Gibbs, S. W. Koch,  Phys. Rev. Lett. {\bf 83}, 2841 (1999); J. P. Prineas, C. Ell, E. S. Lee, G. Khitrova, H. M. Gibbs, S. W. Koch,  Phys. Rev. B {\bf 61}, 13863 (2000);


\bibitem{DeychQW} L. I. Deych, and A. A. Lisyansky,  Phys. Rev. B  {\bf 60}, 4242 (2000).

\bibitem{Lifshitz}  I.M. Lifshitz and A.M. Kosevich, in: {\it Lattice Dynamics} (Benjamin, New York, 1969), p. 53.

\bibitem{Citrinlocal} D. S. Citrin,  Appl. Phys. Lett. {\bf 66}, 994(1995).  

\bibitem{Dereux}  A. Dereux, J.-P. Vigneron, P. Lambin, and A. Lucas, Phys. Rev. B {\bf 38}, 5438 (1988).

\bibitem{Lahlaouti} M. L. H. Lahlaouti, A. Akjouj, B. Djafari-Rouhani, and L. Dobrzynski, Phys. Rev. B {\bf 61}, 2059 (2000).

\bibitem{Andreanireview} L.C. Andreani, ``Optical Transitions, Excitons, and Polaritons in Bulk and Low-Dimensional Semiconductor Structures,'' in: {\it Confined Electrons and Photons} (edited by E. Burstein and C. Weisbuch, Plenum, New York, 1995).

\bibitem{Raikh} A. Yu. Sivachenko, M. E. Raikh, and Z. V. Vardenu, cond-mat/0005396.

\bibitem{Deych}  L. I. Deych and A. A. Lisyansky, Phys. Lett. A {\bf 243}, 156 (1998); L. I. Deych, A. Yamilov, and A. A. Lisyansky, Europhys. Lett. {\bf 46}, 524 (1999).

\bibitem{PRBlocal}  L. I. Deych, A. Yamilov, and A. A. Lisyansky, Phys. Rev. B {\bf 59}, 11339 (1999).

\bibitem{JOSAB}  A. Yamilov, L. I. Deych, and A. A. Lisyansky,  J. Opt. Soc. Amer. B {\bf 17}, 1498 (2000).


\bibitem{Deutsch} I. H. Deutsch, R. J. C. Spreeuw, S. L. Rolston, and W. D. Phillips,  Phys. Rev. A {\bf 52}, 1394 (1995).

\bibitem{LDT} C. Ell, J. Prineas, T. R. Nelson, Jr., S. Park, H. M. Gibbs, G. Khitrova, and S. W. Koch , Phys. Rev. Lett. {\bf 80}, 4795 (1998).

\bibitem{Keldysh} L. V. Keldysh, Superlattices and Microstructure {\bf 4}, 637 (1988).

\bibitem{Ivchenko2}E. L. Ivchenko, Phys. Solid State {\bf 33}, 1344 (1991).

\bibitem{Fano}  U. Fano, Phys. Rev. {\bf 124}, 1866 (1961).

\bibitem{WB}  G. Breit and E. Wigner, Phys. Rev. {\bf 49}, 519 (1936).

\bibitem{Azbel} M. Ya. Azbel, Solid State Comm. {\bf 45}, 527 (1983).








\end{references}
\end{document}